\documentclass[reprint,superscriptaddress,amsmath,amssymb,aps,showpacs,prl]{revtex4-1}

\usepackage{graphicx}
\usepackage{bm}
\usepackage{epsfig}
\usepackage{amssymb}
\usepackage{amsfonts}
\usepackage{braket}
\usepackage{color}

\newcommand{\ee}{\mathrm{e}}
\newcommand{\ii}{\mathrm{i}}
\usepackage{epstopdf}
\usepackage{hyperref}
\usepackage{float}
\usepackage{bibentry}
\usepackage[caption=false]{subfig}

\newcommand{\ba}{\begin{eqnarray}}
\newcommand{\ea}{\end{eqnarray}}
\newcommand{\bd}{\begin{displaymath}}
\newcommand{\ed}{\end{displaymath}}
\renewcommand{\v}[1]{{\bf #1}}
\newcommand{\bpm}{\begin{pmatrix}}
\newcommand{\epm}{\end{pmatrix}}
\newcommand{\nn}{\nonumber \\}

\newcommand{\ua}{\uparrow}
\newcommand{\da}{\downarrow}
\newcommand{\al}{\alpha}
\newcommand{\ga}{\gamma}

\graphicspath{{figures/}}% Put all figures in this directory.

\begin{document}
\title{Featureless Quantum Insulator on the Honeycomb Lattice}
\author{Panjin Kim}
\affiliation{Department of Physics, Sungkyunkwan University, Suwon
  440-746, Korea}
\author{Hyunyong Lee}
\affiliation{Department of Physics, Sungkyunkwan University, Suwon
  440-746, Korea}
\author{Shenghan Jiang}
\affiliation{Department of Physics, Boston College, Chestnut Hill, MA 02467}
\author{Brayden Ware}
\affiliation{Department of Physics, University of California, Santa Barbara, CA 93106-6105, USA}
\author{Chao-Ming Jian}
\affiliation{Department of Physics, Stanford University, Stanford, California 94305, USA}
\author{Michael Zaletel}
\affiliation{Station Q, Microsoft Research, Santa Barbara, CA 93106-6105, USA}
\author{Jung Hoon Han}
\affiliation{Department of Physics, Sungkyunkwan University, Suwon 440-746, Korea}
\author{Ying Ran}
\affiliation{Department of Physics, Boston College, Chestnut Hill, MA 02467}
\date{\today}

\begin{abstract}
We show how to construct fully symmetric, gapped states without topological order on a honeycomb lattice for $S=1/2$ spins using the language of projected entangled pair states\,(PEPS). An explicit example is given for the virtual bond dimension $D=4$. Four distinct classes differing by lattice quantum numbers are found by applying the systematic classification scheme introduced by two of the authors [S. Jiang and Y. Ran, Phys. Rev. B 92, 104414 (2015)].  Lack of topological degeneracy or other conventional forms of symmetry breaking, and the existence of energy gap in the proposed wave functions, are checked by numerical calculations of the entanglement entropy and various correlation functions. Our work provides the first explicit realization of a featureless quantum insulator for spin-1/2 particles on a honeycomb lattice.
\end{abstract}

\maketitle

{\it Introduction} - A modern theme of much interest in condensed matter systems is the classification of possible phases of quantum matter in low dimensions. First noted in the context of quantum Hall physics, it has become clear that different quantum phases are labeled often by their topological characters rather than broken symmetries as in the conventional Ginzburg-Landau paradigm\,\cite{wen07book}. How to define such quantum orders and classify states accordingly in a precise way has intrigued theorists for several decades.

A powerful guide in the classification effort is the ``no-go" theorem such as the celebrated Lieb-Schultz-Mattis theorem in one dimension\,\cite{lieb04} and its higher-dimensional generalizations due, for instance, to Oshikawa\,\cite{oshikawa00} and Hastings\,\cite{hastings04}, stating that lattice spin models having a half-integer spin per unit cell must remain gapless, or if gapped, would either break conventional symmetries or turn into a topological state with fractionalized excitations. Powerful as they are, though, integer spin systems are not covered by these theorems. In one dimension we have some well-established results for integer-spin chains, e.g. $S=1$ Haldane spin chain, saying that the ground state can be both gapped and featureless.

Turning to two dimensions, search for an analogous featureless phase not addressed by the no-go theorems can best proceed by an explicit identification of $S=1$ models on a Bravais lattice\,\cite{jian15}, or $S=1/2$ models on a honeycomb lattice where an even number of sites form a unit cell\,\cite{kimchi13,kimchi15,jian15}. One such construction was given recently with $S=1$ model on a square lattice\,\cite{jian15}, while attempts to construct featureless states for $S=1/2$ spins on a honeycomb lattice has met with partial success so far\,\cite{kimchi13,kimchi15,jian15}. In this paper, we provide an explicit construction of the spin-1/2 wave function on a honeycomb lattice that preserves the full set of lattice symmetries plus time-reversal and SU(2) spin rotation, in addition to being devoid of topological order. We dub such a state featureless quantum insulator, or FQI. Lacking both symmetry breaking and topological order, such states do not permit a straightforward field-theoretic description\,\cite{jian15}. We instead use the recently developed classification scheme of the tensor network wave functions\,\cite{shenghan15,perez10, zhao10, singh10, bauer11, singh11,weichselbaum12, singh12}, in particular the one proposed by two of the authors~\cite{shenghan15}, to identify all possible spin-1/2 FQIs on the honeycomb lattice for a given bond dimension of the tensor network. Intensive numerical check carried out by the authors confirm that the proposed FQI state is indeed devoid of any conventional order, and has the topological entanglement entropy\,\cite{kitaev06,wen06} of zero.

\begin{figure}
  \includegraphics[width=0.4\textwidth]{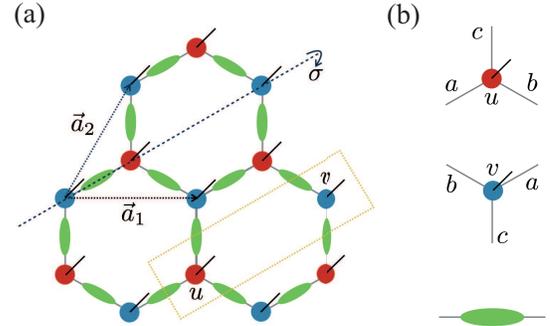}
  \caption{(Color online) Schematic figures of the graphical
    representation of (a) tensor network on honeycomb lattice
    (b) $u$- and $v$-site tensors comprised of three gray legs and upward black
    leg describing Hilbert spaces of three virtual spins
    and a single physical one respectively and bond tensor\,(green
    ellipses) connecting neighboring virtual spins.
    Bravais unit vectors are chosen as $\v a_1=\hat{x}$ and
    $\v a_2=(\hat{x}+\sqrt{3}\hat{y})/2$.}
  \label{fig:schematic}
\end{figure}

{\it Spin-$\frac{1}{2}$ symmetric PEPS on honeycomb lattice} - The honeycomb lattice we work on and various notations for site and bond labels are shown schematically in Fig.\,\ref{fig:schematic}(a). To construct a honeycomb PEPS, we associate every site/bond of the honeycomb lattice with a site/bond tensor. A site tensor is formed by one physical leg which supports physical spin-1/2 degrees of freedom,  and three virtual legs, while a bond tensor is formed by two virtual legs forming a nearest-neighbor bond as shown in Fig.\,\ref{fig:schematic}(b). Since every leg is associated with a specific local Hilbert space of spins, a tensor can be viewed as a quantum state in the Hilbert space of the tensor product of all its leg Hilbert spaces. The physical wave function is obtained by contracting all connected virtual legs of site tensors and bond tensors.

In order to construct a fully symmetric, topologically trivial, and energetically gapped state with spin-1/2 per site on a honeycomb lattice, we rely on the recently developed classification algorithm of quantum phases in terms of PEPS proposed by two of the present authors in Ref. \cite{shenghan15}. Let us briefly review the procedure here. Due to the existence of an enlarged Hilbert space by virtual legs, the mapping from site/bond tensors to physical wave functions are actually many to one. Namely, a wave function which is globally symmetry-preserving can have its constituent site/bond tensors ``gauge-transformed", by acting with an arbitrary invertible matrix $V(s,i)$ on a virtual leg $i$ of the site tensor at $s$ and simultaneously with its inverse matrix $V^{-1}(s,i)$ on the corresponding leg of the bond tensor. There may also exist special gauge transformations that leave every site/bond tensor invariant, up to U(1) phase factor. Those transformations form a group, named the invariant gauge group (IGG), which governs the low energy gauge dynamics of the state as shown in Refs.\,\onlinecite{swingle10,schuch10}.

Let us consider various symmetry operations on PEPS. According to the general remarks above, invariance of the physical wave function $|\psi\rangle$ (up to a U(1) phase) under a specific symmetry operation $g$ implies the following general transformation rules for the site and bond tensors:

\begin{align}
  T^s&=\Theta_g W_g ( g\circ T^s )\notag\\
  T^b&=W_g ( g\circ T^b ) .
  \label{}
\end{align}
We label the site tensor at $s$ as $T^s$, and $T^b$ as the tensor at the bond $b$. $W_g$ is a leg-dependent gauge transformation acting on virtual legs of tensors. Here $g \circ$ can be spin rotation, time reversal or any lattice symmetry operations on a {\it physical} Hilbert space. Symmetry implementation is {\it projective} in the sense that the operation on the physical indices by $g$ can be ``compensated for" by the gauge operations $W_g$ on virtual indices and the U(1) phase factor $\Theta_g$. The site-dependent phase factor $\Theta_g$ also allows us to capture the symmetry quantum numbers of the state $|\psi\rangle$\,\cite{shenghan15}.

Following the framework developed in Ref.\,\onlinecite{shenghan15}, symmetry group operations pertinent to the particular lattice geometry can be cast as a set of algebraic equations. By solving them, one obtains highly constrained forms of all the gauge transformation matrices $W_g$ and $\Theta_g$ associated with the physical symmetric operation $g$. We should mention that IGG will in general enter these algebraic equations, influencing the outcome of the solutions for $W_g$ and $\Theta_g$. In keeping with the spirit of the present paper, which is the search for FQI in the case of the spin-1/2 honeycomb lattice, we set IGG to be trivial, i.e. as an identity element.

Details of the classification procedure for symmetric PEPS with trivial IGG and how to solve for the $W_g$'s and $\Theta_g$'s
are found in the Supplementary Information (SI)\,\cite{si}. We should emphasize that the final expression for the FQI tensors are quite transparent and can be understood without the full knowledge of the classification scheme. In the end, we obtain

\begin{align}
  &W_{T_1}(s,i)=W_{T_2}(s,i)=\mathbb{I},\nn
  &W_{C_6}(u,a/b)=W_{C_6}(v,i)=\mathbb{I},\quad
  W_{C_6}(u,c)=\chi_{C_6},\nn
  %& W_{R_\pi}(u,a/c)=\chi_{C_6},\quad W_{R_\pi}(u,b)=1,\notag\\
 % &W_{R_\pi}(v,a/c)=1,\quad W_{R_\pi}(v,b)=\chi_{C_6};\nn
  &W_{\sigma}(u,a)=W_{\sigma},\quad
  W_\sigma(u,b/c)=\chi_{C_6} W_\sigma,\nn
  &W_\sigma(v,a)=\chi_{C_6}\,\chi_{\sigma C_6}\,W_\sigma,\quad
  W_{\sigma}(v,b/c)=W_\sigma,\nn
  & W_\mathcal{T}(s,i)=W_\mathcal{T}
  =\bigoplus_{k=1}^M(\mathbb{I}_{d_k}\otimes\ee^{\ii\pi S_k^y}),\notag\\
  &W_{\theta\vec{n}}(s,i)=W_{\theta\vec{n}}
  =\bigoplus_{k=1}^M(\mathbb{I}_{d_k}\otimes\ee^{\ii\theta\vec{n}\cdot\vec{S}_k}),\nn
  &\Theta_{T_1}(s)=\Theta_{T_2}(s)=1,\quad
  \Theta_{C_6}(u)=\chi_{C_6},\quad\Theta_{C_6}(v)=1,\notag\\
  %&\Theta_{R_\pi}(u)=1,\quad\Theta_{R_\pi}(v)=\chi_{C_6};\notag\\
  &\Theta_{\sigma}(u)=1,\quad\Theta_{\sigma}(v)
  =\chi_{C_6}\chi_{\sigma C_6},\quad\Theta_\mathcal{T}=1 .
  \label{eq:symmetry_rule_virtual_leg}
\end{align}
Some words on the notations are in order. Two sublattice sites within the unit cell\,(yellow dashed box in Fig.\,\ref{fig:schematic}) are labeled $s = u,\,v$, while $i=a,\,b,\,c$ label the three virtual legs associated with each lattice site as indicated in Fig.\,\ref{fig:schematic}\,(b). Various symmetry operators considered here are translations by lattice vectors $\v a_1$ and $\v a_2$, denoted respectively as $T_1$ and $T_2$, six-fold rotations about the center of the hexagon ($C_6$), mirror reflection $\sigma$ along the dashed line in Fig.\,\ref{fig:schematic}, time-reversal symmetry (${\cal T}$), and spin rotation symmetry. Each virtual leg has the Hilbert space consisting of $M$ different species of spins, each labeled as $\vec{S}_k$, $1\le k \le M$. For each spin $\vec{S}_k$ one further introduces the ``flavor" degeneracy of $d_k$, for a total
virtual Hilbert space dimension $D=\sum_{k=1}^M d_k (2S_k +1)$. To have trivial IGG, we are required to assign only half-integer spins at the virtual legs\,\cite{si}. We further have
\begin{eqnarray}
  W_{\sigma} = \bigoplus_{k=1}^M \left(
    \widetilde{W}_{\sigma}^k
    \otimes \mathbb{I}_{2S_k +1} \right),\label{eq:W-sigma}
\end{eqnarray}
where $\widetilde{W}_\sigma^k$ is a $d_k$-dimensional real matrix satisfying $(\widetilde{W}^k_{\sigma})^2=\mathbb{I}_{d_k}$. Notice that all $W_g$'s in Eq. (\ref{eq:symmetry_rule_virtual_leg}) are translationally invariant (independent of $s$).

\begin{center}
  \begin{table}
    \begin{tabular}{| c || c | c | c | c |}
      \hline
      ( $\chi_{C_6}$, $\chi_{\sigma C_6}$) &$ (+1,+1)$& $(+1,-1)$ &$(-1,+1)$ &
      $(-1,-1)$ \\
      \hline \hline
      $C_6$ & $+1$ & $+1$ &$-1$ &$ -1$ \\ \hline
      $\sigma$ & $+1$ &$ -1$ &$ -1$ & $+1$ \\ \hline
    \end{tabular}
    \caption[Table caption text]{Lattice quantum numbers ($\chi_{C_6}, \chi_{\sigma C_6} )$ for $C_6$ and $\sigma$ operations for four different classes of featureless states.  Sign of the wave function $|\psi\rangle$ changes by the amount shown in the second and the third rows under the $C_6$ and $\sigma$ operations, respectively, for each state characterized by the pair of quantum numbers $(\chi_{C_6}, \chi_{\sigma C_6})$ in the first row. }
    \label{table:lattice_qn}
  \end{table}
\end{center}

According to our analysis\,\cite{si}, there are four different symmetric featureless PEPS classes characterized by $\chi_{C_6}=\pm1$ and $\chi_{\sigma C_6}=\pm1$, regardless of the bond dimension $D$. The PEPS wave functions belonging to different classes indeed can be distinguished by their lattice quantum numbers obtained for the torus geometry with odd number of unit cells, as shown in Table\,\ref{table:lattice_qn}.  For instance, a PEPS wave function with $(\chi_{C_6},\chi_{\sigma C_6}) = (+1,-1)$ is invariant under the $C_6$ operation and gains an extra phase $-1$ under the $\sigma$ operation. After these formal matters, the remaining task is the explicit construction of site and bond tensors and the examination of physical properties for the state obtained from contracting the site/bond tensors. It should be cautioned that, even when the PEPS wave function is seemingly invariant under all symmetry operations, there is a chance that it actually describes a spontaneous symmetry breaking phase. To rule out these possibilities and to ensure that the constructed PEPS state is indeed a symmetric FQI, one should carefully measure the correlation functions for varying system sizes.

We will focus on a particular case where every virtual leg accommodates $n$ copies of spin-1/2's, and the symmetry class where $\chi_{C_6} = \chi_{\sigma C_6} = -1$. While this is not the unique way to derive FQI, the imminent goal of this paper is to  show how to produce {\it an example} of the featureless state for honeycomb spin-1/2's, which is achievable with this particular choice of the symmetry class. One can choose the bond tensor to be the maximally entangled state $T^b =\mathbb{I}_n \otimes \ii\sigma_2$, where $\mathbb{I}_n$ acts on the flavor space and $\ii\sigma_2$ denotes the spin singlet formed by two virtual spin-1/2's. For the site tensor, the most general form of a spin singlet (satisfying spin-rotation symmetry) and Kramers singlet (satisfying time-reversal symmetry) tensor is given by
\begin{align}
  \hat{T}^s =\sum_{\al,\beta,\ga} &\Bigl( \mathcal{C}_{\al\beta\ga}^1 \left( |\uparrow; \downarrow_\al \uparrow_\beta\downarrow_\ga\rangle +|\downarrow; \uparrow_\alpha \downarrow_\beta  \uparrow_\ga\rangle \right)\nn
  & +\mathcal{C}_{\al\beta\ga}^2 \left( |\uparrow; \downarrow_\al \downarrow_\beta\uparrow_\ga\rangle +|\downarrow; \uparrow_\al \uparrow_\beta\downarrow_\ga\rangle \right)\nn
  & + \mathcal{C}_{\al\beta\ga}^3\left( |\uparrow; \uparrow_\al \downarrow_\beta\downarrow_\ga\rangle +|\downarrow; \downarrow_\al \uparrow_\beta\uparrow_\ga\rangle \right) \Bigr) .
\label{eq:local_tensor}
\end{align}
Each element of $\mathcal{C}^i$ is real to preserve the time reversal symmetry, and $\mathcal{C}^1+\mathcal{C}^2+\mathcal{C}^3=0$ due to the SU(2) spin rotation symmetry. The first spin inside the ket separated by the semicolon denotes the physical spin, the other three are virtual spins from each of the three legs for a given site, and $\al,\,\beta,\,\ga$ label the flavor of virtual spins, with $1 \le \al , \beta, \gamma \le n$. For $D=2$ (a single virtual spin-1/2 per leg), there is no PEPS solution satisfying all of lattice symmetries, hence we turn to the simplest non-trivial case with $D=4$ (two flavors of virtual spin-1/2's, $n=2$). Setting $\widetilde{W}_{\sigma} = \sigma_3$ in Eq. (\ref{eq:W-sigma}), we find that, in order to meet the
condition of invariance under the $C_6$ and $\sigma$ symmetries, only the following two independent solutions for the site tensor are possible:
\begin{eqnarray}
  \hat{A}^{(1)} &=& \mathcal{P}\Big(
    2 |\ua; \da_2 \ua_1 \da_2 \rangle - |\ua; \da_1 \ua_2\da_2\rangle
    - |\ua; \da_1 \da_2\ua_2\rangle \nn
    &&~
    + 2 |\da; \ua_2 \da_1 \ua_2 \rangle - |\da; \ua_1 \da_2\ua_2\rangle
    - |\da; \ua_1 \ua_2\da_2\rangle \Big),\nn
  \hat{A}^{(2)} &=& \mathcal{P}\Big(
  | \ua; \da_2 \ua_1 \da_1 \rangle - |\ua; \da_1 \ua_1 \da_2\rangle \nn
  &&~
  +| \da; \ua_2 \da_1 \ua_1 \rangle - |\da; \ua_1 \da_1 \ua_2 \rangle
  \Big). \label{eq:A1A2}
\end{eqnarray}
$\mathcal{P}$ stands for cyclic permutation of the virtual states. The general site tensor consistent with all symmetry requirements can be written as a linear combination $\hat{T}^s = c_1 \hat{A}^{(1)}+c_2 \hat{A}^{(2)}$, with arbitrary real coefficients $c_1,c_2$. We claim that the topologically trivial symmetric PEPS state is obtained from contracting all virtual legs of the site tensors $T^s = c_1 \hat{A}^{(1)}+c_2 \hat{A}^{(2)}$ and bond tensors $T^b$, for appropriate choices of $(c_1, c_2)$.

There are two special cases, $c_1=0$ or $c_2=0$, for which the PEPS wave functions have the emergent U(1) IGG. For $\hat{T}^s=\hat{A}^{(1)}$ in Eq. (\ref{eq:A1A2}), each ket state has two out of the three virtual spins with flavor index equal to 2, and one virtual spin with the flavor index 1. The state made from contracting $\hat{A}^{(1)}$ obviously preserves the flavor quantum number.
A U(1) operation $U(\theta)$, defined by multiplying the (flavor index)=1 virtual spin by $e^{i\theta}$ but not the (flavor index)=2 spin, changes the site tensor by the phase $e^{i \theta}$. These gauge transformations form a U(1) group and result in low-energy fluctuations of U(1) gauge fields, which is known to be confining at long wavelengths in two dimenstions\,\cite{polyakov87}. The same argument shows that the state made out of $\hat{A}^{(2)}$ will likely describe the U(1) spin liquid.

\begin{figure}[htbp]
   \includegraphics[width=0.48\textwidth]{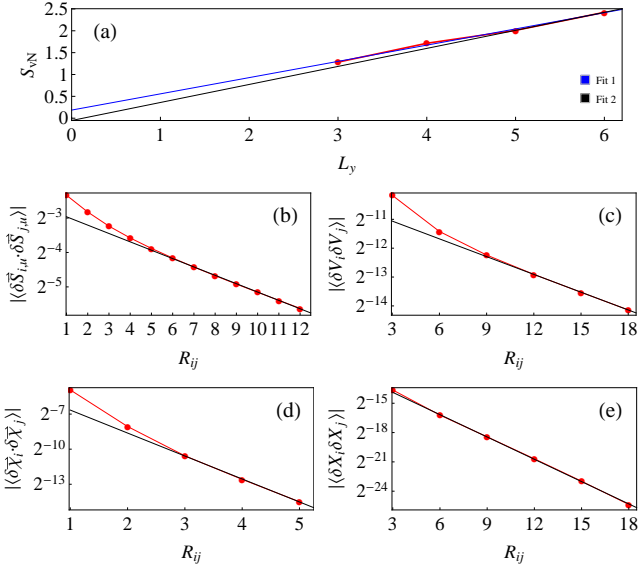}
   \caption{(Color online) Numerical results on spin-1/2 PEPS for $\hat{T}^s = 0.9 \hat{A}^{(1)} + 0.1
     \hat{A}^{(2)}$.
   (a) Entanglement entropy as a function of $L_y$. Two linear fits are shown, based on numerical data at $L_y \in \{3,4,5,6\}$ and $L_y \in \{5,6\}$, respectively.
   (b)-(e) Plots of correlation functions  ($ \delta A = A-\langle A \rangle $) for (b) spin, (c) bond\,($V_i = \v S_{iu}
     \cdot \v S_{iv}$), (d) vector chirality\,($\bm \chi_i = \v S_{iu} \times \v S_{iv}$) and (e) scalar chirality\,($X_i = \v S_{i-1,v} \cdot \v S_{iu} \times \v S_{iv} $) as a function of the distance
     $R_{ij} = |\v x_i - \v x_j|/|\v a_1|$; $\v x_i$ is the
     position vector of the $i$-th unit cell. System size is fixed at $L_x=37$.}
    \label{fig:spin_results}
 \end{figure}

Based on extensive numerical analyses for general $c_2/c_1$ ratios, we conclude that a featureless state without topological order, spontaneous symmetry breaking, or emergent gauge symmetry has been found at around $(c_1 , c_2) = (0.9,0.1)$. First, topological entanglement entropy has been extracted by fitting the calculated entanglement entropy for varying system sizes. As shown in Fig. \ref{fig:spin_results}(a) we find the extrapolated value -0.05 consistent with the absence of topological order. To evaluate the entanglement entropy, we imposed the periodic boundary condition along the $\vec{a}_2$-direction of our ansatz wave function and employed the boundary theory of PEPS\,\cite{cirac11, wahl14}.

 \begin{figure}[htbp]
   \includegraphics[width=0.5\textwidth]{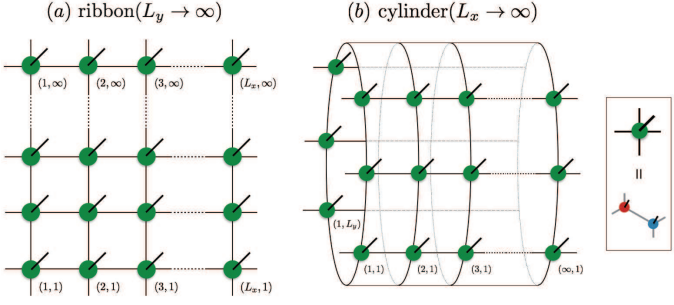}
   \caption{(Color online) Schematic figure of the lattice geometry used to evaluate (a) correlators and (b) entanglement entropy. A green dot denotes one unit cell, formed by $u$\,(red) and $v$\,(blue) sites of the honeycomb lattice.}
    \label{fig:geometry}
 \end{figure}

Correlation functions were measured for spin, bond, vector spin chirality, and scalar spin chirality, in order to determine the gapped nature of the state and to check for the absence of symmetry breaking. For convenience in numerical calculation, a unit cell was redefined so that $u$- and $v$-site tensors are directly connected within a unit cell such that $\v x_{i,u} -\, \v x_{i,v} = (\v a_2 - 2 \v a_1)/3$, where $\v x_{i,u(v)}$ is the position vector of $u(v)$-tensor at the $i$-th unit cell.  Expectation values of local operators $\langle \psi | O_i | \psi\rangle$ and correlators $\langle \psi | O_i O_j| \psi \rangle $ are obtained by employing the MPS-MPO compression method~\cite{verstraete04,schollwock11,orus14} to approximately contract a given tensor network. Two-site variational compression is adopted~\cite{schollwock11} where the initial ansatz is given by the zip-up algorithm~\cite{miles10}. The reduced bond dimension $d$, which is equal to the number of singular values kept during the compression process of MPS, is such that the truncated singular value is less than $10^{-7}$. To minimize the numerical instability caused by lower and upper boundaries of the tensor network wave function, compression is repeated until the convergence of MPS $|\Psi_n \rangle \simeq \mathcal{O} |\Psi_{n-1} \rangle$ is achieved, where $|\Psi_n \rangle$ is the $n$th stage MPS and $\mathcal{O}$ is the MPO. Number of unit cells in the $\v a_1$ direction is 37.

\begin{figure}[htbp]
  \includegraphics[width=0.48\textwidth]{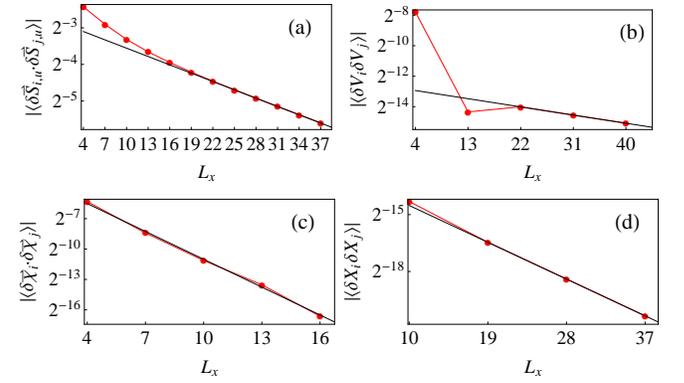}
  \caption{(Color online) (a) Spin, (b) bond, (c) vector chirality, and (d) scalar chirality correlations as functions of $L_x$, at the separation distance equal to $L_x/2$. Exponential decays are observed in all correlators. }
  \label{fig:scaling_size}
\end{figure}

As one can see in Fig.\,\ref{fig:spin_results}, all correlators decay exponentially in the fixed system size $L_x =37$. We double-checked the exponential decay by varying $L_x$ and fixing the position of operators $O_i$ and $O_j$ to be at $L_x/3$ and $2L_x/3$, respectively. Exponential decays are clearly observed for all correlators as a function of $L_x$ in Fig. \ref{fig:scaling_size}. Lattice and spin rotation symmetries are numerically confirmed as well. Finally, we extracted the correlation lengths as a function of reduced bond dimension $d$ and found them to saturate for large enough $d$~\,\cite{si}.
Based on such overwhelming body of evidences, we conclude that our ansatz PEPS is a topologically trivial, fully symmetric and gapped quantum state or FQI.

We also measured the entanglement entropy for the U(1) states ($c_1=0$ or $c_2=0$). In both cases, it depends on the number of flavor 1 states and flavor 2 states we put on the boundary virtual legs. Physically, we can interpret flavor states as U(1) gauge charges. For boundary conditions with different total flavor numbers, we end up with wave functions supporting different number of electric field lines along the length of cylinder. These states are orthogonal to each other, and in general give different entanglement entropies.\\

{\it Conclusion} - We have identified an exemplary state of a featureless quantum insulator on the honeycomb lattice of spin-1/2's, based on the methodical search scheme developed in Ref.\,\cite{shenghan15}. Four distinct classes have been identified as a result of our search. We propose a state whose physical properties are consistent with FQI. Compared to previous works on FQI's~\cite{kimchi13,kimchi15,jian15}, the present method offers a much more systematic way to classify tensor network states consistent with symmetry and topological constraints. The gapped liquid phase we constructed is intrinsically strongly interacting, as there is no way to adiabatically connect them to free electronic states. These results may thus be relevant for correlated electronic materials on the honeycomb lattice. For instance, evidences for a putative spin liquid ground state have been reported for Ba$_3$CuSb$_2$O$_9$\,\cite{shanavas14,smerald14}, in which the spin-1/2 Cu may form a honeycomb lattice\,\cite{nakatsuji12}.

{\it Acknowledgements} -  SHJ and YR are supported by the Alfred P. Sloan fellowship and National Science Foundation under Grant No. DMR-1151440. HYL is supported by the NRF grant (No.2015R1D1A1A01059296).
CMJ is supported by the David and Lucile Packard foundation and National Science Foundation under Grant No. NSF PHY11- 25915.

\bibliographystyle{apsrev}
\bibliography{reference.bib}

\begin{thebibliography}{30}
\expandafter\ifx\csname natexlab\endcsname\relax\def\natexlab#1{#1}\fi
\expandafter\ifx\csname bibnamefont\endcsname\relax
  \def\bibnamefont#1{#1}\fi
\expandafter\ifx\csname bibfnamefont\endcsname\relax
  \def\bibfnamefont#1{#1}\fi
\expandafter\ifx\csname citenamefont\endcsname\relax
  \def\citenamefont#1{#1}\fi
\expandafter\ifx\csname url\endcsname\relax
  \def\url#1{\texttt{#1}}\fi
\expandafter\ifx\csname urlprefix\endcsname\relax\def\urlprefix{URL }\fi
\providecommand{\bibinfo}[2]{#2}
\providecommand{\eprint}[2][]{\url{#2}}

\bibitem[{\citenamefont{Wen}(2007)}]{wen07book}
\bibinfo{author}{\bibfnamefont{X.}~\bibnamefont{Wen}},
  \emph{\bibinfo{title}{Quantum Field Theory of Many-Body Systems: From the
  Origin of Sound to an Origin of Light and Electrons}}, Oxford Graduate Texts
  (\bibinfo{publisher}{OUP Oxford}, \bibinfo{year}{2007}), ISBN
  \bibinfo{isbn}{9780199227259},
  \urlprefix\url{https://books.google.com/books?id=1fxpPgAACAAJ}.

\bibitem[{\citenamefont{Lieb et~al.}(2004)\citenamefont{Lieb, Schultz, and
  Mattis}}]{lieb04}
\bibinfo{author}{\bibfnamefont{E.}~\bibnamefont{Lieb}},
  \bibinfo{author}{\bibfnamefont{T.}~\bibnamefont{Schultz}}, \bibnamefont{and}
  \bibinfo{author}{\bibfnamefont{D.}~\bibnamefont{Mattis}}, in
  \emph{\bibinfo{booktitle}{Condensed Matter Physics and Exactly Soluble
  Models}} (\bibinfo{publisher}{Springer}, \bibinfo{year}{2004}), pp.
  \bibinfo{pages}{543--601}.

\bibitem[{\citenamefont{Oshikawa}(2000)}]{oshikawa00}
\bibinfo{author}{\bibfnamefont{M.}~\bibnamefont{Oshikawa}},
  \bibinfo{journal}{Phys. Rev. Lett.} \textbf{\bibinfo{volume}{84}},
  \bibinfo{pages}{1535} (\bibinfo{year}{2000}),
  \urlprefix\url{http://link.aps.org/doi/10.1103/PhysRevLett.84.1535}.

\bibitem[{\citenamefont{Hastings}(2005)}]{hastings04}
\bibinfo{author}{\bibfnamefont{M.}~\bibnamefont{Hastings}},
  \bibinfo{journal}{EPL (Europhysics Letters)} \textbf{\bibinfo{volume}{70}},
  \bibinfo{pages}{824} (\bibinfo{year}{2005}).

\bibitem[{\citenamefont{Jian and Zaletel}(2016)}]{jian15}
\bibinfo{author}{\bibfnamefont{C.-M.} \bibnamefont{Jian}} \bibnamefont{and}
  \bibinfo{author}{\bibfnamefont{M.}~\bibnamefont{Zaletel}},
  \bibinfo{journal}{Phys. Rev. B} \textbf{\bibinfo{volume}{93}},
  \bibinfo{pages}{035114} (\bibinfo{year}{2016}),
  \urlprefix\url{http://link.aps.org/doi/10.1103/PhysRevB.93.035114}.

\bibitem[{\citenamefont{Kimchi et~al.}(2013)\citenamefont{Kimchi, Parameswaran,
  Turner, Wang, and Vishwanath}}]{kimchi13}
\bibinfo{author}{\bibfnamefont{I.}~\bibnamefont{Kimchi}},
  \bibinfo{author}{\bibfnamefont{S.}~\bibnamefont{Parameswaran}},
  \bibinfo{author}{\bibfnamefont{A.~M.} \bibnamefont{Turner}},
  \bibinfo{author}{\bibfnamefont{F.}~\bibnamefont{Wang}}, \bibnamefont{and}
  \bibinfo{author}{\bibfnamefont{A.}~\bibnamefont{Vishwanath}},
  \bibinfo{journal}{Proceedings of the National Academy of Sciences}
  \textbf{\bibinfo{volume}{110}}, \bibinfo{pages}{16378}
  (\bibinfo{year}{2013}).

\bibitem[{\citenamefont{Ware et~al.}(2015)\citenamefont{Ware, Kimchi,
  Parameswaran, and Bauer}}]{kimchi15}
\bibinfo{author}{\bibfnamefont{B.}~\bibnamefont{Ware}},
  \bibinfo{author}{\bibfnamefont{I.}~\bibnamefont{Kimchi}},
  \bibinfo{author}{\bibfnamefont{S.~A.} \bibnamefont{Parameswaran}},
  \bibnamefont{and} \bibinfo{author}{\bibfnamefont{B.}~\bibnamefont{Bauer}},
  \bibinfo{journal}{Phys. Rev. B} \textbf{\bibinfo{volume}{92}},
  \bibinfo{pages}{195105} (\bibinfo{year}{2015}),
  \urlprefix\url{http://link.aps.org/doi/10.1103/PhysRevB.92.195105}.

\bibitem[{\citenamefont{Jiang and Ran}(2015)}]{shenghan15}
\bibinfo{author}{\bibfnamefont{S.}~\bibnamefont{Jiang}} \bibnamefont{and}
  \bibinfo{author}{\bibfnamefont{Y.}~\bibnamefont{Ran}},
  \bibinfo{journal}{Phys. Rev. B} \textbf{\bibinfo{volume}{92}},
  \bibinfo{pages}{104414} (\bibinfo{year}{2015}),
  \urlprefix\url{http://link.aps.org/doi/10.1103/PhysRevB.92.104414}.

\bibitem[{\citenamefont{P{\' e}rez-Garc{\' i}a et~al.}(2010)\citenamefont{P{\'
  e}rez-Garc{\' i}a, Sanz, Gonz{\' a}lez-Guill{\' e}n, Wolf, and
  Cirac}}]{perez10}
\bibinfo{author}{\bibfnamefont{D.}~\bibnamefont{P{\' e}rez-Garc{\' i}a}},
  \bibinfo{author}{\bibfnamefont{M.}~\bibnamefont{Sanz}},
  \bibinfo{author}{\bibfnamefont{C.~E.} \bibnamefont{Gonz{\' a}lez-Guill{\'
  e}n}}, \bibinfo{author}{\bibfnamefont{M.~M.} \bibnamefont{Wolf}},
  \bibnamefont{and} \bibinfo{author}{\bibfnamefont{J.~I.} \bibnamefont{Cirac}},
  \bibinfo{journal}{New Journal of Physics} \textbf{\bibinfo{volume}{12}},
  \bibinfo{pages}{025010} (\bibinfo{year}{2010}),
  \urlprefix\url{http://stacks.iop.org/1367-2630/12/i=2/a=025010}.

\bibitem[{\citenamefont{Zhao et~al.}(2010)\citenamefont{Zhao, Xie, Chen, Wei,
  Cai, and Xiang}}]{zhao10}
\bibinfo{author}{\bibfnamefont{H.~H.} \bibnamefont{Zhao}},
  \bibinfo{author}{\bibfnamefont{Z.~Y.} \bibnamefont{Xie}},
  \bibinfo{author}{\bibfnamefont{Q.~N.} \bibnamefont{Chen}},
  \bibinfo{author}{\bibfnamefont{Z.~C.} \bibnamefont{Wei}},
  \bibinfo{author}{\bibfnamefont{J.~W.} \bibnamefont{Cai}}, \bibnamefont{and}
  \bibinfo{author}{\bibfnamefont{T.}~\bibnamefont{Xiang}},
  \bibinfo{journal}{Phys. Rev. B} \textbf{\bibinfo{volume}{81}},
  \bibinfo{pages}{174411} (\bibinfo{year}{2010}),
  \urlprefix\url{http://link.aps.org/doi/10.1103/PhysRevB.81.174411}.

\bibitem[{\citenamefont{Singh et~al.}(2010)\citenamefont{Singh, Pfeifer, and
  Vidal}}]{singh10}
\bibinfo{author}{\bibfnamefont{S.}~\bibnamefont{Singh}},
  \bibinfo{author}{\bibfnamefont{R.~N.~C.} \bibnamefont{Pfeifer}},
  \bibnamefont{and} \bibinfo{author}{\bibfnamefont{G.}~\bibnamefont{Vidal}},
  \bibinfo{journal}{Phys. Rev. A} \textbf{\bibinfo{volume}{82}},
  \bibinfo{pages}{050301} (\bibinfo{year}{2010}),
  \urlprefix\url{http://link.aps.org/doi/10.1103/PhysRevA.82.050301}.

\bibitem[{\citenamefont{Bauer et~al.}(2011)\citenamefont{Bauer, Corboz, Or\'us,
  and Troyer}}]{bauer11}
\bibinfo{author}{\bibfnamefont{B.}~\bibnamefont{Bauer}},
  \bibinfo{author}{\bibfnamefont{P.}~\bibnamefont{Corboz}},
  \bibinfo{author}{\bibfnamefont{R.}~\bibnamefont{Or\'us}}, \bibnamefont{and}
  \bibinfo{author}{\bibfnamefont{M.}~\bibnamefont{Troyer}},
  \bibinfo{journal}{Phys. Rev. B} \textbf{\bibinfo{volume}{83}},
  \bibinfo{pages}{125106} (\bibinfo{year}{2011}),
  \urlprefix\url{http://link.aps.org/doi/10.1103/PhysRevB.83.125106}.

\bibitem[{\citenamefont{Singh et~al.}(2011)\citenamefont{Singh, Pfeifer, and
  Vidal}}]{singh11}
\bibinfo{author}{\bibfnamefont{S.}~\bibnamefont{Singh}},
  \bibinfo{author}{\bibfnamefont{R.~N.~C.} \bibnamefont{Pfeifer}},
  \bibnamefont{and} \bibinfo{author}{\bibfnamefont{G.}~\bibnamefont{Vidal}},
  \bibinfo{journal}{Phys. Rev. B} \textbf{\bibinfo{volume}{83}},
  \bibinfo{pages}{115125} (\bibinfo{year}{2011}),
  \urlprefix\url{http://link.aps.org/doi/10.1103/PhysRevB.83.115125}.

\bibitem[{\citenamefont{Weichselbaum}(2012)}]{weichselbaum12}
\bibinfo{author}{\bibfnamefont{A.}~\bibnamefont{Weichselbaum}},
  \bibinfo{journal}{Annals of Physics} \textbf{\bibinfo{volume}{327}},
  \bibinfo{pages}{2972 } (\bibinfo{year}{2012}), ISSN
  \bibinfo{issn}{0003-4916},
  \urlprefix\url{http://www.sciencedirect.com/science/article/pii/S0003491612001121}.

\bibitem[{\citenamefont{Singh and Vidal}(2012)}]{singh12}
\bibinfo{author}{\bibfnamefont{S.}~\bibnamefont{Singh}} \bibnamefont{and}
  \bibinfo{author}{\bibfnamefont{G.}~\bibnamefont{Vidal}},
  \bibinfo{journal}{Phys. Rev. B} \textbf{\bibinfo{volume}{86}},
  \bibinfo{pages}{195114} (\bibinfo{year}{2012}),
  \urlprefix\url{http://link.aps.org/doi/10.1103/PhysRevB.86.195114}.

\bibitem[{\citenamefont{Kitaev and Preskill}(2006)}]{kitaev06}
\bibinfo{author}{\bibfnamefont{A.}~\bibnamefont{Kitaev}} \bibnamefont{and}
  \bibinfo{author}{\bibfnamefont{J.}~\bibnamefont{Preskill}},
  \bibinfo{journal}{Phys. Rev. Lett.} \textbf{\bibinfo{volume}{96}},
  \bibinfo{pages}{110404} (\bibinfo{year}{2006}),
  \urlprefix\url{http://link.aps.org/doi/10.1103/PhysRevLett.96.110404}.

\bibitem[{\citenamefont{Levin and Wen}(2006)}]{wen06}
\bibinfo{author}{\bibfnamefont{M.}~\bibnamefont{Levin}} \bibnamefont{and}
  \bibinfo{author}{\bibfnamefont{X.-G.} \bibnamefont{Wen}},
  \bibinfo{journal}{Phys. Rev. Lett.} \textbf{\bibinfo{volume}{96}},
  \bibinfo{pages}{110405} (\bibinfo{year}{2006}),
  \urlprefix\url{http://link.aps.org/doi/10.1103/PhysRevLett.96.110405}.

\bibitem[{\citenamefont{Swingle and Wen}(2010)}]{swingle10}
\bibinfo{author}{\bibfnamefont{B.}~\bibnamefont{Swingle}} \bibnamefont{and}
  \bibinfo{author}{\bibfnamefont{X.-G.} \bibnamefont{Wen}},
  \bibinfo{journal}{arXiv preprint arXiv:1001.4517}  (\bibinfo{year}{2010}).

\bibitem[{\citenamefont{Schuch et~al.}(2010)\citenamefont{Schuch, Cirac, and
  P{\' e}rez-Garc{\' i}a}}]{schuch10}
\bibinfo{author}{\bibfnamefont{N.}~\bibnamefont{Schuch}},
  \bibinfo{author}{\bibfnamefont{I.}~\bibnamefont{Cirac}}, \bibnamefont{and}
  \bibinfo{author}{\bibfnamefont{D.}~\bibnamefont{P{\' e}rez-Garc{\' i}a}},
  \bibinfo{journal}{Annals of Physics} \textbf{\bibinfo{volume}{325}},
  \bibinfo{pages}{2153 } (\bibinfo{year}{2010}), ISSN
  \bibinfo{issn}{0003-4916},
  \urlprefix\url{http://www.sciencedirect.com/science/article/pii/S0003491610000990}.

\bibitem[{\citenamefont{Kim et~al.}(2016)\citenamefont{Kim, Lee, Jiang, Ware,
  Jian, Zaletel, Han, and Ran}}]{si}
\bibinfo{author}{\bibfnamefont{P.}~\bibnamefont{Kim}},
  \bibinfo{author}{\bibfnamefont{H.}~\bibnamefont{Lee}},
  \bibinfo{author}{\bibfnamefont{S.}~\bibnamefont{Jiang}},
  \bibinfo{author}{\bibfnamefont{B.}~\bibnamefont{Ware}},
  \bibinfo{author}{\bibfnamefont{C.-M.} \bibnamefont{Jian}},
  \bibinfo{author}{\bibfnamefont{M.}~\bibnamefont{Zaletel}},
  \bibinfo{author}{\bibfnamefont{J.}~\bibnamefont{Han}}, \bibnamefont{and}
  \bibinfo{author}{\bibfnamefont{Y.}~\bibnamefont{Ran}},
  \bibinfo{journal}{Supplementary Information}  (\bibinfo{year}{2016}).

\bibitem[{\citenamefont{Polyakov}(1987)}]{polyakov87}
\bibinfo{author}{\bibfnamefont{A.~M.} \bibnamefont{Polyakov}},
  \emph{\bibinfo{title}{Gauge fields and strings}}, vol. \bibinfo{volume}{140}
  (\bibinfo{publisher}{Harwood academic publishers Chur},
  \bibinfo{year}{1987}).

\bibitem[{\citenamefont{Cirac et~al.}(2011)\citenamefont{Cirac, Poilblanc,
  Schuch, and Verstraete}}]{cirac11}
\bibinfo{author}{\bibfnamefont{J.~I.} \bibnamefont{Cirac}},
  \bibinfo{author}{\bibfnamefont{D.}~\bibnamefont{Poilblanc}},
  \bibinfo{author}{\bibfnamefont{N.}~\bibnamefont{Schuch}}, \bibnamefont{and}
  \bibinfo{author}{\bibfnamefont{F.}~\bibnamefont{Verstraete}},
  \bibinfo{journal}{Phys. Rev. B} \textbf{\bibinfo{volume}{83}},
  \bibinfo{pages}{245134} (\bibinfo{year}{2011}),
  \urlprefix\url{http://link.aps.org/doi/10.1103/PhysRevB.83.245134}.

\bibitem[{\citenamefont{Wahl et~al.}(2014)\citenamefont{Wahl, Ha\ss{}ler, Tu,
  Cirac, and Schuch}}]{wahl14}
\bibinfo{author}{\bibfnamefont{T.~B.} \bibnamefont{Wahl}},
  \bibinfo{author}{\bibfnamefont{S.~T.} \bibnamefont{Ha\ss{}ler}},
  \bibinfo{author}{\bibfnamefont{H.-H.} \bibnamefont{Tu}},
  \bibinfo{author}{\bibfnamefont{J.~I.} \bibnamefont{Cirac}}, \bibnamefont{and}
  \bibinfo{author}{\bibfnamefont{N.}~\bibnamefont{Schuch}},
  \bibinfo{journal}{Phys. Rev. B} \textbf{\bibinfo{volume}{90}},
  \bibinfo{pages}{115133} (\bibinfo{year}{2014}),
  \urlprefix\url{http://link.aps.org/doi/10.1103/PhysRevB.90.115133}.

\bibitem[{\citenamefont{Verstraete and Cirac}(2004)}]{verstraete04}
\bibinfo{author}{\bibfnamefont{F.}~\bibnamefont{Verstraete}} \bibnamefont{and}
  \bibinfo{author}{\bibfnamefont{J.~I.} \bibnamefont{Cirac}},
  \bibinfo{journal}{arXiv preprint cond-mat/0407066}  (\bibinfo{year}{2004}).

\bibitem[{\citenamefont{Schollwöck}(2011)}]{schollwock11}
\bibinfo{author}{\bibfnamefont{U.}~\bibnamefont{Schollwöck}},
  \bibinfo{journal}{Annals of Physics} \textbf{\bibinfo{volume}{326}},
  \bibinfo{pages}{96 } (\bibinfo{year}{2011}), ISSN \bibinfo{issn}{0003-4916},
  \bibinfo{note}{january 2011 Special Issue},
  \urlprefix\url{http://www.sciencedirect.com/science/article/pii/S0003491610001752}.

\bibitem[{\citenamefont{Or{\' u}s}(2014)}]{orus14}
\bibinfo{author}{\bibfnamefont{R.}~\bibnamefont{Or{\' u}s}},
  \bibinfo{journal}{Annals of Physics} \textbf{\bibinfo{volume}{349}},
  \bibinfo{pages}{117 } (\bibinfo{year}{2014}), ISSN \bibinfo{issn}{0003-4916},
  \urlprefix\url{http://www.sciencedirect.com/science/article/pii/S0003491614001596}.

\bibitem[{\citenamefont{Stoudenmire and White}(2010)}]{miles10}
\bibinfo{author}{\bibfnamefont{E.}~\bibnamefont{Stoudenmire}} \bibnamefont{and}
  \bibinfo{author}{\bibfnamefont{S.~R.} \bibnamefont{White}},
  \bibinfo{journal}{New Journal of Physics} \textbf{\bibinfo{volume}{12}},
  \bibinfo{pages}{055026} (\bibinfo{year}{2010}).

\bibitem[{\citenamefont{Shanavas et~al.}(2014)\citenamefont{Shanavas,
  Popovi\ifmmode~\acute{c}\else \'{c}\fi{}, and Satpathy}}]{shanavas14}
\bibinfo{author}{\bibfnamefont{K.~V.} \bibnamefont{Shanavas}},
  \bibinfo{author}{\bibfnamefont{Z.~S.}
  \bibnamefont{Popovi\ifmmode~\acute{c}\else \'{c}\fi{}}}, \bibnamefont{and}
  \bibinfo{author}{\bibfnamefont{S.}~\bibnamefont{Satpathy}},
  \bibinfo{journal}{Phys. Rev. B} \textbf{\bibinfo{volume}{89}},
  \bibinfo{pages}{085130} (\bibinfo{year}{2014}),
  \urlprefix\url{http://link.aps.org/doi/10.1103/PhysRevB.89.085130}.

\bibitem[{\citenamefont{Smerald and Mila}(2014)}]{smerald14}
\bibinfo{author}{\bibfnamefont{A.}~\bibnamefont{Smerald}} \bibnamefont{and}
  \bibinfo{author}{\bibfnamefont{F.}~\bibnamefont{Mila}},
  \bibinfo{journal}{Phys. Rev. B} \textbf{\bibinfo{volume}{90}},
  \bibinfo{pages}{094422} (\bibinfo{year}{2014}),
  \urlprefix\url{http://link.aps.org/doi/10.1103/PhysRevB.90.094422}.

\bibitem[{\citenamefont{Nakatsuji et~al.}(2012)\citenamefont{Nakatsuji, Kuga,
  Kimura, Satake, Katayama, Nishibori, Sawa, Ishii, Hagiwara, Bridges
  et~al.}}]{nakatsuji12}
\bibinfo{author}{\bibfnamefont{S.}~\bibnamefont{Nakatsuji}},
  \bibinfo{author}{\bibfnamefont{K.}~\bibnamefont{Kuga}},
  \bibinfo{author}{\bibfnamefont{K.}~\bibnamefont{Kimura}},
  \bibinfo{author}{\bibfnamefont{R.}~\bibnamefont{Satake}},
  \bibinfo{author}{\bibfnamefont{N.}~\bibnamefont{Katayama}},
  \bibinfo{author}{\bibfnamefont{E.}~\bibnamefont{Nishibori}},
  \bibinfo{author}{\bibfnamefont{H.}~\bibnamefont{Sawa}},
  \bibinfo{author}{\bibfnamefont{R.}~\bibnamefont{Ishii}},
  \bibinfo{author}{\bibfnamefont{M.}~\bibnamefont{Hagiwara}},
  \bibinfo{author}{\bibfnamefont{F.}~\bibnamefont{Bridges}},
  \bibnamefont{et~al.}, \bibinfo{journal}{Science}
  \textbf{\bibinfo{volume}{336}}, \bibinfo{pages}{559} (\bibinfo{year}{2012}),
  \eprint{http://www.sciencemag.org/content/336/6081/559.full.pdf},
  \urlprefix\url{http://www.sciencemag.org/content/336/6081/559.abstract}.

\end{thebibliography}

\end{document}